\documentclass[sigconf]{acmart}

\usepackage{multirow}

\AtBeginDocument{%
  }

\setcopyright{rightsretained}
\copyrightyear{2023}
\acmYear{2023}
\acmDOI{}

\acmConference[MuRS @ RecSys '23]{}{2023}{Singapore}

\acmBooktitle{Proceedings of MuRS Workshop, RecSys 2023} 
\acmPrice{}
\acmISBN{}

\begin{document}

\title{Popularity Degradation Bias in Local Music Recommendation}

\author{April Trainor}
\email{atrainor@ithaca.edu}
\orcid{0000-0003-0364-9019}
\affiliation{%
  \institution{Ithaca College}
  \streetaddress{953 Danby Rd.}
  \city{Ithaca}
  \state{New York}
  \country{USA}
  \postcode{14850}
}

\author{Douglas R. Turnbull}
\email{dturnbull@ithaca.edu}
\orcid{0009-0001-7252-1855}
\affiliation{%
  \institution{Ithaca College}
  \streetaddress{953 Danby Rd.}
  \city{Ithaca}
  \state{New York}
  \country{USA}
  \postcode{14850}
}

\renewcommand{\shortauthors}{Trainor and Turnbull}

\begin{abstract}
In this paper, we study the effect of popularity degradation bias in the context of local music recommendations. Specifically, we examine how accurate two top-performing recommendation algorithms, Weight Relevance Matrix Factorization (WRMF) and Multinomial Variational Autoencoder (Mult-VAE), are at recommending artists as a function of artist popularity. We find that both algorithms improve recommendation performance for more popular artists and, as such, exhibit popularity degradation bias. While both algorithms produce a similar level of performance for more popular artists, Mult-VAE shows better relative performance for less popular artists. This suggests that this algorithm should be preferred for local (long-tail) music artist recommendation.

\end{abstract}

\received{6 Aug 2023}
\received[accepted]{28 Aug 2023}

\maketitle

\section{Introduction}
Researchers have studied local music scenes and how they can have a positive impact on their surrounding communities. These benefits include the promotion of social cohesion,  improved emotional well-being, and increased economic activity \cite{baker2020community, titan2015austin,oakes2011conceptualizing, zhou2022towards}. 
To this end, we have developed a music artist and live event recommendation system called \emph{Localify.org} in an attempt to help people better connect with their local music scenes \cite{turnbull2023localify}. Our primary goal is to recommend relevant local artists similar to the artists that a user already listens to.

When using Localify, a user first selects a set of favorite \emph{seed} artists and a location. Our recommendation algorithm then ranks a set of \emph{local candidate} artists. These local candidate artists either originate from or have an upcoming show near the user's location. To contextualize each recommended artist, we associate each recommendation with one or more of the user's seed artists and we provide relevant genre tags.  Users can then interact with an artist recommendation by reading the artist's biography, listening to a song clip, seeing a list of upcoming (local) events, and seeing a list of similar artists. Finally, we generate a personalized playlist of all the artist recommendations on the user's streaming music service (Spotify or Apple Music.)  By exposing users to relevant local artists, we hope that our users then support these artists by listening to their music, attending their live events, and sharing their music with others. 

One unique aspect of Localify is that the vast majority of local artists are relatively unpopular and reside in the long-tail of the artist popularity distribution \cite{anderson04, Schedl2022Music}. As a result, we are interested in developing a recommender system that is particularly good at recommending long-tail artists.

In this paper, we compare the performance of two standard recommender system algorithms, Weighted Relevance Matrix Factorization (WRMF) by Hu et al. \cite{hu2008collaborative} and Multinomial Variational Autoencoders (Mult-VAE) by Liang et al. \cite{liang2018variational} across different artist popularity ranges. These models were chosen due to their competitive performance shown in recent studies \cite{dacrema2019we, rendle2020neural}. Our goal is to measure the effect of \emph{popularity-related degradation bias} \cite{turnbull2023localify, steck2011item} on music recommendation.

Another challenge when developing our music recommender system is our initial lack of access to user-item preference data. Instead, we can use public music information APIs from Spotify and Apple to collect a large amount of artist similarity data. We show how we can adapt both WRMF and Mult-VAE to use an artist-artist similarity matrix to make relevant music artist recommendations. 

We also use these APIs to collect popularity scores for each artist.  This popularity information allows us to both tune the hyperparameters of our recommendation algorithms as well as  measure popularity-related degradation. In Section \ref{sec:exp_design} we describe this experimental design which is motivated by the Localify user experience.

\section{Artist Origin and Music Event Data }\label{sec:music_data}

As of August 2023, we collected information for 1.5 million artists and artist origin (``hometown'', ``born'', ``formed'') information for 41,526 of these artists. We also collected information for 5.8 million events involving 243,076 artists. However, due to disruptions caused to live music events by the COVID-19 pandemic, for this paper, we are going to consider the subset of 1.1 million recent or upcoming events starting from January 1, 2022. This set of events involves 99,276 unique artists. Our event information was collected using the Bandsintown API \footnote{https://artists.bandsintown.com/support/public-api} and aggregating Google events. We collected artist origin information from sites including Bandsintown, Apple Music, Wikipedia, Rate Your Music\footnote{https://rateyourmusic.com/}, in addition to contributions by Localify.org users.

\begin{table}[h!]
\begin{tabular}{|l|c|cccc|}
\hline
\multirow{2}{*}{}                                                              & \multirow{2}{*}{\begin{tabular}[c]{@{}c@{}}Number\\ Artists\end{tabular}} & \multicolumn{4}{c|}{Spotify Popularity}                                                                       \\ \cline{3-6} 
                                                                               &                                                                           & \multicolumn{1}{l|}{25\%} & \multicolumn{1}{l|}{50\%} & \multicolumn{1}{l|}{75\%} & \multicolumn{1}{l|}{95\%} \\ \hline
\begin{tabular}[c]{@{}l@{}}Artists with \\ known origins\end{tabular}          & 41,526                                                                    & \multicolumn{1}{c|}{5}    & \multicolumn{1}{c|}{22}   & \multicolumn{1}{c|}{35}   & 54                        \\ \hline
\begin{tabular}[c]{@{}l@{}}Artists with events \\ in 2022 to 2023\end{tabular} & 99,276                                                                    & \multicolumn{1}{c|}{3}    & \multicolumn{1}{c|}{16}   & \multicolumn{1}{c|}{33}   & 52                        \\ \hline
\end{tabular}

\caption{ \label{table:artist_pop_stats} Spotify popularity percentile statistics for artists with origin information and recent or current live events. }
\end{table}

In Table \ref{table:artist_pop_stats}, we examine the relationship between Spotify popularity (on a scale from 0-100)  and these two different types of \emph{local} artists: artists with origin information and artists with events. We rank each of these sets of artists and report Spotify popularity at the 25th, 50th (i.e., median), 75th, and 95th percentiles. We note that the majority of these artists are relatively obscure (median Spotify popularity of 22 for artists with origin information and 16 for artists with recent events). This suggests that it will be important to optimize our recommendation algorithms for low-popularity artists if our goal is to provide the best possible recommendations for local artists and events.

%
%
%

%
%

%

%
%
%

%

%

%

%

\section{Bootstrapping  Recommendation with Artist Similarity Data}
Recommender systems typically utilize interaction or preference data between users and items in order to recommend new items. These data are often represented as a user-item matrix. This type of system is common due to the ready availability of implicit or explicit feedback in commercial settings, e.g. 'user $a$ watched item $b$', or 'user $a$ gave item $b$ a thumbs down'. When creating a new recommendation system, in general, we do not have access to a large  amount of user preference data. 

As such, we utilize public APIs from music technology companies (Spotify\footnote{https://developer.spotify.com/documentation/web-api}, Apple Music\footnote{https://developer.apple.com/documentation/applemusicapi/}) instead. These APIs provide us with a set of similar artists for a given artist, and Spotify's additionally provides us with a relative measure of their popularity (0-100). By employing this approach, we can construct a large artist-artist similarity matrix through iterative API queries. The process involves retrieving information for a particular artist, then proceeding to obtain data on their similar artists, and subsequently repeating the process using snowball sampling \cite{goodman1961snowball}. For each artist, we also collect their popularity score as well as a list of their top genres from Spotify, which will be used later in our evaluation procedure.

For our experiments, we consider two complementary recommendation algorithms: WRMF and Mult-VAE. WRMF is a top-performing (linear) matrix factorization algorithm designed for modeling implicit feedback \cite{rendle2020neural}. Mult-VAE is a (non-linear) deep neural network model that has shown highly competitive performance when compared to other deep neural network and matrix factorization models in recent studies \cite{dacrema2019we}. In the following two subsections, we describe how we adapt these recommendation algorithms to work with an artist-artist similarity matrix.

\subsection{ Weighted Relevance Matrix Factorization (WRMF)}

Weighted Relevance Matrix Factorization (WRMF) \cite{hu2008collaborative} is a collaborative filtering algorithm that was explicitly designed for implicit user feedback. Typically, matrix factorization is used to decompose the user-item matrix such that each user and item is embedded into a $k$-dimensional space. Since the user-item matrix is very sparse and represented with binary values (the user has or has not listened to the artist), a weight $\alpha$  is included to make all of the observed interactions count more in the objective function.  Finally, ridge regression (l2-norm) is used for regularization to prevent overfitting.  The amount of regularization in the objective function is controlled by the hyperparameter $\lambda$.  

For WRMF, we can directly use our artist-artist similarity matrix in place of the user-item matrix:  each artist is embedded into a $k$-dimensional space. We can then represent a user as a sparse  $|\mathcal{A}|$-dimensional vector with a value of 1 for each dimension corresponding to each of the user's seed artists.   We then embed this sparse user vector into the $k$-dimensional artist similarity space. Finally, we calculate and sort the distances between this user vector and each of the candidate artist vectors. 

For our experiments, we use the Implicit Python library\footnote{https://github.com/benfred/implicit} implementation of WRFM which uses alternating least squares (ALS) optimization. We conduct a grid search over the three hyperparameters 
($k$, $\lambda$, $\alpha$) 
and pick values for them based on the best performance when considering low-popularity artists with Spotify popularity between 20-24. This mimics performance on the task of local artist recommendation which is described in Section \ref{sec:exp_design}.   Our final hyperparameter settings were set to $k=128$,  $\lambda=0.1$, and $\alpha=15$ when training the WRMF model for our experiment.

\subsection{Multinomial Variational Autoencoder (Mult-VAE)}

Multinomial Variational Autoencoder (Mult-VAE) \cite{liang2018variational} is a probabilistic generative model used for collaborative filtering in recommender systems. It uses two parts, an encoder to map user-item interactions into a low-dimensional latent space, and a decoder which maps the latent space into a probability distribution of all items. The encoder and decoder are represented as a multilayer perceptron (MLP), and implementations typically have one or more hidden layers between the encoder and decoder to learn additional item features. Non-linear activation functions are used between layers, e.g. tanh or sigmoid. The input to Mult-VAE is a vector of binary interaction data, and the output is the probability of each item being liked. As such, it has no internal representation of individual user preferences, unlike WRMF. This allows predictions to be performed directly via forward passes through the MLP from a set of a user's previous interactions without requiring additional optimization computation at the time when we recommend artists to a user.

As with WRMF, we can directly use our artist-artist matrix  in place of a user-item matrix when training a Mult-VAE model if we treat each artist as a user and construct a vector out of that artist's similar artists. During training, batches of artists are selected and are forward-propagated through the MLP.  The loss is calculated and propagated backwards through the model, updating weights. Overfitting is controlled through dropout \cite{srivastava2014dropout} at the input layer with probability 0.2. We use the PyTorch Python library to implement Mult-VAE. In our implementation, we use Mean Squared Error (MSE) as our loss function, and Adam \cite{kingma2015adam} as our optimization algorithm. Our network utilizes one hidden layer, and the dimension of the bottleneck layer is 200. We use tanh as our activation function between layers. For training, we perform mini-batch updates using 250 artists at a time.

\section{Experimental Design for Long-Tail Recommendation}\label{sec:exp_design}

As discussed in Section \ref{sec:music_data}, most of the local artists that are recommended by Localify are low-popularity. To this end, we would like to evaluate the relative performance of the recommendation algorithms when recommending these more obscure artists. To do this, we propose that we simulate both users and local music scenes. That is, we represent a simulated user as an individual who likes ten mainstream seed artists from each of two music genres. Similarly, a simulated music scene might have ten obscure candidate artists from each of eight genres.  The goal of the recommendation algorithm is thus, given a simulated user's 20 seed artists, rank the 80 local candidate artists such that 20 local candidate artists who are associated with one of the user's two preferred genres are at the top of the ranking.    

\begin{table}[h!]
\begin{tabular}{|c c c c|}
\hline
rock       & jazz       & punk       & reggae    \\ 
electronic & metal      & indie r\&b & metalcore \\ 
pop        & indie      & latin      & classical \\ 
folk       & country    & dubstep    & indie pop \\ 
rap        & tech house & norteno    & house     \\ \hline
\end{tabular}
\caption{\label{tab:genres} List of 20 genres used in our experiments}
\end{table}

For each trial, the eight music scene genres are randomly chosen from a list of 20 common genres that are listed in Table \ref{tab:genres}. These 20 genres selected are the product of a greedy algorithm maximizing artist coverage over the union of the 41K artists with origin information and the 99K artists with recent or upcoming events. That is, we selected the genre associated with the most artists in this subset, removed all artists with that genre from the subset, and repeated this process 20 times.

Our experimental design is as follows:

\vspace{2mm}
    For 16 Spotify Popularly Ranges $\in$  \{0-4, 5-9, ..., 75-79\}
    \vspace{0mm}
    \begin{enumerate}
        \item Repeat 100 Times
        \begin{enumerate}
            \item Pick 8 \emph{Music Scene Genres} from a set of 20 common music genres
            \item Pick 2 \emph{User Seed Genres} from the 8 Music Scene Genres
            \item Pick 80 random \emph{candidate artists} to represent the music scene, with 10 artists per music scene genre such that each artist's popularity is within the given range
            \item Pick 20 random \emph{user preference artists}, with 10 artists per user seed genre such that each artist is randomly sampled from the top 100 most popular artists for that genre
            \item for each recommendation algorithm (WRMF, Mult-VAE)
            \begin{enumerate}
                \item{rank the 80 candidate artists given the user preference artists}
                \item reveal which of the 20 candidate artists share one of the two user seed genres
                \item calculate the AUC score for the ranking for artists
            \end{enumerate}

        \end{enumerate}
        \item calculate the average AUC score for each recommendation algorithm for each of the popularity ranges
    \end{enumerate}

We use the standard area under the Receiver Operation Characteristic curve (AUC) metrics \cite{schutze2008introduction} to measure the quality of the candidate artist recommendation rankings. The AUC score would be 1 if the recommendation algorithm places all 20 of the relevant candidate artists at the top of the ranking. The AUC score is expected to be 0.5 if the recommendation algorithm were to randomly shuffle the 80 candidate artists. 
We note that many genres like \emph{reggae}, \emph{classical}, and \emph{jazz}, do not have associated artists with Spotify popularity scores above 80 which is why we are unable to consistently compute AUC scores above this threshold. In rare cases in which there are not 10 artists associated with a genre for a given popularity range, we select as many artists as we can that meet the criteria. However, the expected value of the AUC evaluation metric is invariant to the proportion of relevant to non-relevant items in a ranked list which allows us to fairly average AUC scores across multiple trials.

\section{Results}

\begin{figure}
    \centering
    \includegraphics[width=3.3in]{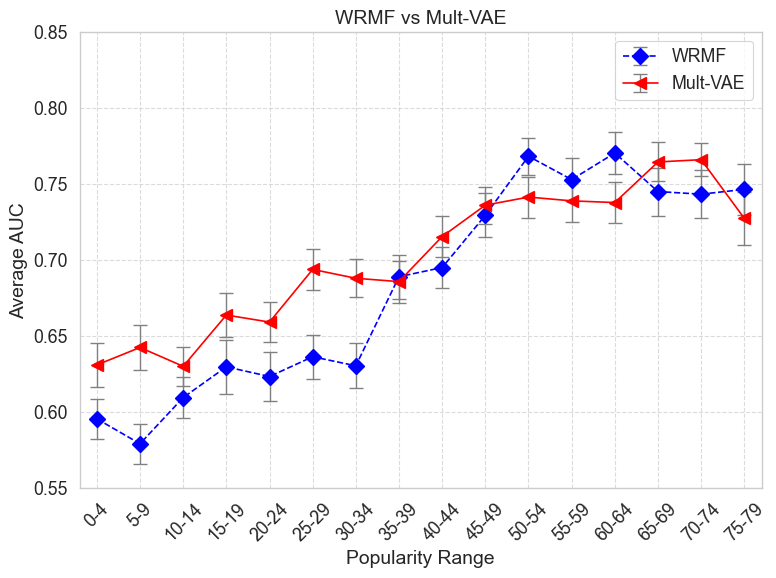}
    \caption{Plot of Average AUC with standard error bars across different Spotify artist popularity ranges. We see that both recommendation algorithms, WRMF and Mult-VAE, show a decrease in performance as a function of decreasing artist popularity and thus reveal popularity degradation bias. Mult-VAE shows better relative performance for artists with lower popularity.  }
    \label{fig:pop_degrade}
\end{figure}

The results for our experiment are shown in Figure \ref{fig:pop_degrade}. We see a clear upward trend in average AUC as a function of Spotify artist popularity for both WRMF and Mult-VAE. However, the slope of the improvement is larger for WMRF suggesting that WMRF is more susceptible to popularity degradation bias. If we were to divide up the plot into approximately two parts, Mult-VAE seems to outperform WMRF at low levels of artist popularity (0-34) while both perform about the same for higher levels of popularity (35-80). 

For our Localify use case which involves mostly recommending less popular local artists, this means we would prefer to use Mult-VAE if we were to only consider recommendation performance. We should note that WRMF takes much less computational time and resources to train, which is important as we are constantly retraining in a production setting. We also note that WRMF and Mult-VAE are both fast to evaluate when making recommendations for a user in real-time.

\section{Discussion}
Our goal for this research was to explore the magnitude of popularity degradation bias for two commonly used recommender system algorithms. When looking specifically at Mult-VAE we do observe this bias in our experimental results but the difference is perhaps reasonable in a production setting. That is, users tend to be more critical of recommendations when they are familiar, and more accepting of a recommendation when the artist is unknown to them. As a result, a small overall difference in average AUC of 0.67 for the less popular artists (20-25) to 0.76 for the most popular artist (75-80) may not be perceptible by a typical user when using a music recommendation system like Localify.org. 

Future work will involve both exploring other top-performing recommendation algorithms \cite{dacrema2019we} (e.g., SLIM \cite{ning2011slim}, Neural MF \cite{he2017neural}) as well as modifying existing algorithms to better optimize for recommending less popular artists. For example, the loss function of the WRMF algorithm is optimized \emph{per-interaction}, and since more popular items tend to have more user-item interactions, the algorithm is thus optimized to fit more popular items. However, if we were to modify this objective function to be optimized \emph{per-item} instead, we might not observe as much popularity degradation bias.

\section{Acknowledgments}
This research was supported in part by NSF Award IIS-1901330. All content represents the opinion of the authors, which is not necessarily shared or endorsed by their respective employers and/or sponsors.

\bibliographystyle{ACM-Reference-Format}
\bibliography{Workshop_RecSys23}

\appendix

\end{document}